\documentclass[preprint]{revtex4}
\usepackage{graphicx}
\usepackage{amssymb}
\newcommand{\be}{\begin{eqnarray}}
\newcommand{\ee}{\end{eqnarray}}
\newcommand{\ATslash}{\rlap{\,/} A_T}
\newcommand{\delslash}{\rlap{/} \partial}

\begin{document}

\title{Twist Three Distribution e(x) : Sum Rules and Equation of 
Motion Relations}
\author{\bf Asmita Mukherjee$^a$}
\affiliation{$^a$ Department of Physics,\\
Indian Institute of Technology Bombay, Powai, Mumbai 400076,
India.}
\date{\today}

\begin{abstract}
We investigate the twist three distribution function $e(x)$ in light-front
Hamiltonian perturbation theory. In light-front gauge, by eliminating the
constrained field, we find a mass term, an intrinsic transverse momentum
dependent term and a 'genuine twist three' quark-gluon interaction term in
the operator. The equation of motion relation, moment relation and the
sum rules are satisfied for a quark at one loop. We compare the results with
other model calculations. 
\end{abstract}
\maketitle
\noindent
{\bf Introduction}
\vspace{.2cm}

\noindent
Twist three parton distribution $e(x)$ has not been explored broadly in the
literature so far. $e(x)$ is spin independent and chiral odd. Therefore it
is difficult to measure it experimentally as it can combine only with
another chiral odd object. It was shown in \cite{jaffe} that $e(x)$
contributes to the unpolarized Drell-Yan process at the level of twist four. 
$e(x)$ enters together with the chirally odd Collin's fragmentation 
function $H_1^\perp$ in the azimuthal asymmetry in semi-inclusive deep
inelastic scattering (DIS) of longitudinally polarized electrons off
unpolarized nucleons. This asymmetry was measured \cite{exp1,exp2,exp3},       
however, apart from $e(x)$, several other distribution and fragmentation
functions also appear in this asymmetry. In \cite{radici}, another
possibility of measuring $e(x)$ is proposed through two hadron production in
polarized semi-inclusive DIS; where $e(x)$ is coupled to a two hadron
fragmentation function.  
The first calculation of $e(x)$ was done in
\cite{jaffe}, in MIT bag model. Further it was calculated in chiral quark
soliton model \cite{chiral1,chiral2}, spectator model \cite{spec} and 
perturbative one loop model \cite{burkardt}. The $Q^2$
evolution of $e(x)$ has been calculated  in \cite{belitsky}. A nice review
of the general properties of $e(x)$ and the various model calculations can
be found in \cite{eqm}. 
In chiral quark soliton model, a $\delta(x)$ singularity was observed
in $e(x)$, which was later found also in a perturbative calculation
\cite{burkardt}.
However, no such $\delta(x)$ term was found in the spectator model as well
as in MIT bag model. It is to be noted that as $x = 0$ cannot be
experimentally achieved, no direct experimental verification of the
$\delta(x)$ contribution is possible. Another interesting aspect of $e(x)$ 
is that the first
moment of the flavor singlet combination of $e(x)$ is related to the pion
nucleon sigma term $\sigma_{\pi  N}$ \cite{jaffe}. The flavor non-singlet part of $e(x)$
satisfies a sum rule connecting it to the hadronic mass difference between
neutron and proton \cite{chiral1}.  The first moment of $e(x)$ obeys the sum rule    
\be
\int_{-1}^1~ dx~ e(x)={1\over 2  M} \langle P,S \mid {\bar{\psi}}(0) 
\psi(0) \mid P, S \rangle.
\label{sum1}
\ee
The second moment of $e(x)$ obeys another sum rule \cite{jaffe}:
\be
\int_{-1}^1 ~dx~ x~ e(x)={m\over  M} N_q;
\label{sum2}
\ee
where $m$ is the mass of the quark and $M$ is the mass of the proton. $N_q$ is the number of 
quarks of a given flavor. These sum rules are
not satisfied in the bag model as the QCD equation of motion is modified in the
bag. They are not satisfied in the spectator model either. In QCD equation
of motion method \cite{eqm}, the first sum rule is saturated by the $\delta(x)$
contribution only, whereas in chiral quark soliton model, only  a part of the
contribution comes from the $\delta(x)$ term. In \cite{burkardt}, $e(x)$ has
been calculated for a quark dressed with  a gluon at one loop in QCD in light
front gauge. Starting from the Feynman diagram  in $A^+=0$ gauge, $k^-$ was
integrated out to get to perform the calculation in light-front time-ordered
method. A $\delta(x)$ term was found which was shown to be related  to the
$k^+=0$ modes. 

The disagreement between different model calculations for $e(x)$ makes it an
interesting object for further study. In this work, we calculate $e(x)$ for
a quark at one loop in QCD using light front Hamiltonian perturbation theory
in $A^+=0$ gauge. Instead of using the Feynman diagrams, we expand the state
in Fock space in terms of multiparton light-front wavefunctions. 
The advantage is that
these wave functions are Lorentz boost invariant, so we can truncate the
Fock space expansion to a few particle sector in a boost invariant way. 
The two particle light-front wave functions (LFWFs) can be calculated
analytically for a quark at one loop using the light-front Hamiltonian. 
The partons are on-mass shell objects having non-vanishing transverse momenta.   
They can be called field theoretic partons. The distribution functions can
be calculated at the scale $Q^2$ using the multiparton LFWFs. To
$O(\alpha_s)$ this is an exact calculation.  
Another interesting aspect is the nature of ultraviolet (UV)  divergence 
in $e(x)$. Being a twist three object, $e(x)$ contains a 'bad' component 
of quark field. As  a result, it is expected to have a different UV     
divergence as compared to the logarithmic divergence in twist two
distributions $f_1(x)$. In the following, we present our calculation.

\vspace{.2cm}
\noindent
{\bf e(x) for a quark at one loop}
\vspace{.2cm}

\noindent
The twist three distribution $e(x)$ is given by \cite{jaffe} :  
\be
e(x)={P^+\over M} \int {dy^-\over  8 \pi}e^{{i\over 2} x P^+ y^-} \langle
P,S\mid {\bar{\psi}}(0)\psi(y^-) \mid P,S \rangle.
\label{ex}
\ee
We introduce the projection operators $ \Lambda^\pm={1\over 2} \gamma_0
\gamma_\pm$ and project out the light-front good and bad components of
the quark field : 
\be
\psi^{(\pm)}=\Lambda^\pm \psi
\ee
In light-front gauge, $A^+=0$, the 'bad' component, $\psi^{(-)}$ is constrained,
and the equation of constraint is given by \cite{two}
\be
\psi^-(y^-)={1\over i\partial^+} (i \alpha^\perp \cdot \partial^\perp + 
g \alpha^\perp \cdot A^\perp + \beta m) \psi^{(+)}(y^-);
\ee
where the operator ${1\over \partial^+}$ is defined as \cite{two}
\be
{1\over \partial^+} f(x^-)={1\over 4} \int_{-\infty}^{\infty} dy^-
\epsilon(x^--y^-) f(y^-)
\label{pp}
\ee
The antisymmetric step function is given by 
\be
\epsilon(x^-)=-{i \over \pi} {\cal{P}} \int {d \omega\over \omega} e^{{i \over
2} \omega x^-}
\ee
${\cal{P}}$ denotes the principal value. For the dynamical field 
The operator in Eq. (\ref{ex}), can be written as
\be
O_e={\bar{\psi}}(0) \psi(y^-)=\psi^{(-)\dagger}(0) \gamma_0 \psi^{(+)}(y^-)+
\psi^{(+)\dagger}(0) \gamma_0 \psi^{(-)}(y^-).
\ee
Using the constraint equation for $\psi^{(-)}$ this can be written as:
\be
O_e=O_m+O_g+O_k;
\ee
where
\be
O_m=m \psi^{(+)\dagger}(0) \Big [ (-\stackrel{\leftarrow}{1\over i
\partial^+})+ (\stackrel{\rightarrow} {1\over i \partial^+})\Big ] 
\psi^{(+)}(y^-);
\ee
\be
O_k=\psi^{(+) \dagger} (0) \Big [ \stackrel{\rightarrow}{{\delslash^\perp}
\over \partial^+}-\stackrel{\leftarrow}{{\delslash^\perp}\over \partial^+} 
\Big ] \psi^{(+)}(y^-);
\ee
\be
O_g= g \psi^{(+) \dagger} (0) \Big [ \ATslash (\stackrel{\leftarrow}
{1\over i \partial^+})+ (\stackrel{\rightarrow}{1\over i \partial^+}) \ATslash
\Big ]\psi^{(+)}(y^-).
\ee
$O_m$ is the mass term, $O_k$ is the transverse momentum dependent term and
$O_g$ is the explicit quark-gluon interaction term, also called the
`genuine twist three' term.

For $\psi^+$ we use two component formalism \cite{two}
\be
\psi^+ = \bordermatrix{&  \cr
              &\xi \cr
               & 0}
\ee 
The two component field $\xi(y)$ has the Fock space expansion
\be
\xi(y)=\sum_\lambda \chi_\lambda \int {dk^+ d^2k^\perp\over
 2 {(2 \pi)}^3 \sqrt{k^+}} \Big [
b^\dagger_\lambda (k)e^{i k y }+ d_{-\lambda}(k) e^{-i k y} \Big ];
\ee
with
\be
\chi_{\uparrow} = \bordermatrix{&  \cr
              &1 \cr
               & 0}, ~~~~~~~~~~
\chi_{\downarrow} = \bordermatrix{&  \cr
              &0 \cr
               & 1}.
\ee 


For a dressed quark state of momentum $P$ and helicity $\sigma$:
\begin{eqnarray}
\mid P, \sigma \rangle && = \phi_1 b^\dagger(P,\sigma) \mid 0 \rangle
\nonumber \\  
&& + \sum_{\sigma_1,\lambda_2} \int
{dk_1^+ d^2k_1^\perp \over \sqrt{2 (2 \pi)^3 k_1^+}}
\int 
{dk_2^+ d^2k_2^\perp \over \sqrt{2 (2 \pi)^3 k_2^+}}
\sqrt{2 (2 \pi)^3 P^+} \delta^3(P-k_1-k_2) \nonumber \\
&& ~~~~~\phi_2(P,\sigma \mid k_1, \sigma_1; k_2 , \lambda_2) b^\dagger(k_1,
\sigma_1) a^\dagger(k_2, \lambda_2) \mid 0 \rangle.
\label{eq2}    
\end{eqnarray} 

Here $a^\dagger$ and $b^\dagger$ are bare gluon and quark
creation operators respectively and $\phi_1$ and $\phi_2$ are the
multiparton wave functions. They are the probability amplitudes to find one
bare quark and one quark plus gluon inside the dressed quark state
respectively. 
We introduce Jacobi momenta $x_i$,~${q_i}^\perp$ such that $\sum_i x_i=1$ and
$\sum_i {q_i}^\perp=0$.  They are defined as
\be
x_i={k_i^+\over P^+}, ~~~~~~q_i^\perp=k_i^\perp-x_i P^\perp.
\ee
Also, we introduce the wave functions,
\be
\psi_1=\phi_1, ~~~~~~~~~~~\psi_2(x_i,q_i^\perp)= {\sqrt {P^+}} \phi_2
(k_i^+,{k_i}^\perp);
\ee
which are independent of the total transverse momentum $P^\perp$ of the
state and are boost invariant.
The state is normalized as,   
\be
\langle P',\lambda'\mid P,\lambda \rangle = 2(2\pi)^3
P^+\delta_{\lambda,\lambda'} \delta(P^+-{P'}^+)\delta^2(P^\perp-P'^\perp).
\label{norm}
\ee
The two particle wave function depends on the helicities of the electron and
photon. Using the eigenvalue equation for the light-cone Hamiltonian, this  
can be written as \cite{rajen},
\be
\psi^\sigma_{2\sigma_1,\lambda}(x,q^\perp)&=& {x(1-x)\over
(q^\perp)^2+m^2 (1-x)^2}
{1\over {\sqrt {(1-x)}}} {g\over
{\sqrt {2(2\pi)^3}}} T^a \chi^\dagger_{\sigma_1}\Big [- 2 {q^\perp\over
{1-x}}-{{\tilde \sigma^\perp}\cdot q^\perp\over x} {\tilde \sigma^\perp}
\nonumber\\&&~~~~~~~~~~~~~~~~~~
+i m{\tilde \sigma}^\perp {(1-x)\over x}\Big ]\chi_\sigma
\epsilon^{\perp *}_\lambda \psi_1.
\label{psi2}
\ee
$m$ is the bare mass of the quark, $\tilde \sigma_1=\sigma_2$, $\tilde
\sigma_2=-\sigma_1$. 
$\psi_1$ actually gives the normalization of the state \cite{rajen}:
\be
{\mid \psi_1 \mid}^2=1-{\alpha_s\over {2 \pi}} C_f
\int_\epsilon^{1-\epsilon} dx {{1+x^2}\over {1-x}}~log
{Q^2\over \mu^2},
\label{c5nq}
\ee
to order $\alpha_s$.  Here $\epsilon$ is a small cutoff on $x$. We have
taken the cutoff on the transverse momenta to be $Q$.
This gives the large scale of the process. The above expression 
is derived using Eqs (\ref{norm}), (\ref{eq2}) and
(\ref{psi2}). In the above expression, we have neglected subleading finite
pieces. $\mu$ is a small scale separating hard and soft dynamics such that
$(q^\perp)^2 \ge \mu^2$. For a dressed quark, we get,
\be
x~e_k(x,Q^2)=0;
\ee
\be
x~e_g(x,Q^2)&=& {m\over M} {\alpha_s\over {2 \pi}}~ C_f ~ log{Q^2 \over
\mu^2} \Big [ {x\over 2} \delta(1-x)-1+x \Big ];
\ee
\be 
x~e_m(x,Q^2) &=& {m \over M} 
\Big [ \delta(1-x)+{\alpha_s\over {2 \pi}}~ C_f ~ log{Q^2 \over
\mu^2} \Big \{ {1+x^2\over 1-x}-\delta(1-x) \int_{\epsilon}^{1-\epsilon} dy
{1+y^2\over 1-y} \Big \} \Big ];
\ee 
where $x~e_k(x,Q^2)$, $x~e_g(x,Q^2$ and $x~e_m(x,Q^2)$ are contributions
from $O_k$,$O_g$ and $O_m$ respectively.
In the above, we have used the normalization condition Eq. (\ref{c5nq}). One
can write
\be
{1+x^2\over 1-x}-\delta(1-x) \int_{\epsilon}^{1-\epsilon} dy
{1+y^2\over 1-y}={1+x^2\over 1-x}_+ + {3\over 2} \delta(1-x).
\ee
The plus prescription is defined in the usual way, that is $\int_0^1 dx
{f(x)\over {1-x}_+}=\int_0^1 dx {f(x)-f(1)\over 1-x}$. 
For a dressed quark state, $M$ is the renormalized mass of the quark. 
The bare mass $m$ of the quark is given in terms of the renormalized mass
\cite{gT}:
\be
m=M \Big ( 1-{3 \alpha_s\over 4 \pi} C_f log{Q^2 \over \mu^2} \Big ).     
\ee
Using this, we can write
\be
x e(x,Q^2)= \delta(1-x)+ {\alpha_s\over 2 \pi} C_f log{Q^2 \over \mu^2} 
\Big [{2 x\over {1-x}_+} +{1\over 2} x \delta(1-x)\Big ].
\ee
In the above result, the divergence at $x
\rightarrow 1$ gets canceled by the contribution from the normalization of
the state and we get the plus prescription. The above result agrees with
\cite{burkardt}. Note that there is no 
$\delta(x)$ contribution in $x e(x)$.  However, as shown in 
\cite{burkardt}, $\delta(x)$ contribution in $e(x)$ is 
related to the $k^+=0$ modes (zero modes) in light-front gauge. By choosing 
the  prescription given  by Eq. (\ref{pp}) we are avoiding the zero modes.   
$x e(x) $ is zero if we take the quark mass to be zero.

\vspace{.2cm}
\noindent
{\bf Sum rules and equation of motion relation}
\vspace{.2cm}

\noindent

$x e(x)$ can be related to the twist two unpolarized quark 
distribution through the equation of motion relation \cite{piet}:
\be
x e(x) = x \tilde e(x)+{m\over M} f_1(x)
\label{eqm}
\ee
Where $f_1(x)$ is the twist two unpolarized distribution function :
\be
f_1(x)= \int {dy^-\over  8 \pi}e^{{i\over 2} x P^+ y^-} \langle
P,S\mid {\bar{\psi}}(0) \gamma^+ \psi(y^-) \mid P,S \rangle.
\label{f1}
\ee
Note that Eq. (\ref{eqm}) is unaffected by the presence of the gauge link in
the definition of the parton distributions. In the above relation, we have 
suppressed the scale dependence. $\tilde e(x) $ is the genuine
twist three quark-gluon interaction part of $e(x)$. For a dressed quark,
\be
f_1(x, Q^2)=\delta(1-x)+{\alpha_s\over 2 \pi} C_f log{Q^2 \over \mu^2} \Big
[ {1+x^2\over {1-x}_+}+{3\over 2} \delta(1-x)\Big ]  
\ee
So the equation of motion relation is satisfied with $x~ \tilde e(x,Q^2)= 
{\alpha_s \over 2 \pi}~ C_f~ log {Q^2\over \mu^2} \Big [ 
{1\over 2} x \delta(1-x)-1+x \Big ]$.
The first moment of $e_m(x)$  and $e_g(x)$ are  given by
\be
\int_{\epsilon}^1~ e_m(x,Q^2)~ dx = 1+{\alpha_s\over 2 \pi}~ C_f ~
log{Q^2 \over \mu^2}~ (-\log \epsilon -1);
\ee
\be
\int_{\epsilon}^1~ e_g(x,Q^2)~ dx = {\alpha_s\over 2 \pi}~ C_f ~
log{Q^2 \over \mu^2}~(log \epsilon +{3\over 2}).
\ee
Each part has divergence as $x \rightarrow 0$. However, their total
contribution is free of divergence :
\be
\int_{\epsilon}^1~ e (x,Q^2)~ dx &=& \int_{\epsilon}^1~ (e_m (x,Q^2)+e_g(x,Q^2) 
~ dx =1+ {\alpha_s\over 2 \pi}~ C_f ~
log{Q^2 \over \mu^2}~{1\over 2} \nonumber\\&& ={1\over 2  M} 
\langle P,S \mid {\bar{\psi}}(0) \psi(0) \mid P, S \rangle. 
\ee 
So the sum  rule in Eq. (\ref{sum1}) is satisfied. However, our $x$ 
region is limited by
$0$ and $1$ as this is physically allowed. Note that this implies that the sum rule is
not saturated by a $\delta(x)$ contribution, unlike what was concluded in
\cite{eqm}. The possibility of a divergence
in this relation was observed in \cite{jaffe}. The second moment of
$e(x)$ becomes
\be
\int_0^1~ x~ e(x, Q^2) dx = 1-{\alpha_s\over 2 \pi}~ C_f~ 
log{Q^2 \over \mu^2}~ {3\over 2}= {m\over M};
\label{second}
\ee 
with 
\be
\int_0^1 dx~ x~ \tilde e(x,Q^2)= \int_0^1~dx~\Big [ {x\over 2}~
\delta(1-x)-1+x \Big ]=  0;
\ee
The {\it  rhs} of Eq. (\ref{second}) vanishes in the chiral limit $m =0$. Note that this result
agrees with \cite{burkardt}, the $\delta(x)$ present there in $e(x)$ does
not contribute to this sum rule.
The relation for the $n$-th moment of $e(x)$ defined by ${[e]}_n=\int_0^1
dx~x^{n-1}~e(x)$ can be written as \cite{moment}:
\be 
{[e]}_n = {[\tilde e]}_n+{m\over M}{[f_1]}_{n-1}.
\label{moment}
\ee
The $n$-th moment is calculated using the expression above :
\be
{[e]}_n&=& 1+ {\alpha_s\over 2 \pi}~ C_f~ 
log{Q^2 \over \mu^2}\int_0^1~ dx ~x^{n-2}~ \Big [{2x\over 1-x}_++{1\over 2} x
\delta(1-x) \Big ] \nonumber\\&&
= 1+ {\alpha_s\over 2 \pi}~ C_f~ log{Q^2 \over \mu^2}\Big [ -2
\sum_{j=1}^{n-1}{1\over j}+{1\over 2}\Big ];
\ee
where we have used $\int_0^1 ~ dx ~ x^{n-1}~{1\over 1-x}_+=-\sum_{j=1}^{n-1}
{1\over j}$.
On the {\it rhs},  
\be
{[\tilde e]}_n &=& {\alpha_s\over 2 \pi}~ C_f~ 
log{Q^2 \over \mu^2} \int_0^1 dx x^{n-2}\Big [ {x\over 2} \delta(1-x)-1+x\Big ] \nonumber\\&&
={\alpha_s\over 2 \pi}~ C_f~  log{Q^2 \over \mu^2}
\Big [{1\over 2}-{1\over n-1}+{1\over n}\Big ];
\ee 
\be
{1\over M}{[m f_1]}_{n-1} &=& 1+{\alpha_s\over 2 \pi}~ C_f~ 
log {Q^2 \over \mu^2} \int_0^1~ dx ~x^{n-2}~ \Big [{1+x^2\over 1-x}_+ + {3\over
2}-{3\over 2} \Big ] \nonumber\\&&
= {\alpha_s\over 2 \pi}~ C_f~ 
log {Q^2 \over \mu^2} \Big [-2 \sum_{j=1}^{n-1}{1\over j}
-{1\over n}+{1\over n-1}\Big ];
\ee
So the moment relation Eq. (\ref{moment}) is satisfied.

To summarize, in this Letter, we investigate the twist three distribution
function $e(x,Q^2)$ for a massive quark at one loop 
using light-front Hamiltonian perturbation theory in light-front gauge.
By expressing the operator in terms of dynamical fields, we find three terms
in the operator ; quark mass term , intrinsic transverse momentum dependent
term and an explicit quark-gluon interaction term. The intrinsic transverse
momentum dependent part does not give contribution. The equation of motion
relation directly relates $x e(x,Q^2)$ to the twist two unpolarized distribution
function $f_1(x,Q^2)$. The mass of the quark plays a vital role here. 
The first moment relation for $e(x,Q^2)$ is satisfied  without a $\delta(x)$
term in the distribution function. Contribution from the mass term as well
as the quark-gluon interaction term diverge as $x \rightarrow 0$, however
their total contribution does not diverge in this limit. The
second moment vanishes in the chiral limit. The logarithmic divergence of
$e(x,Q^2)$ gives the scale dependence.

{\bf{Acknowledgment}}
We thank Daniel Boer and Piet Mulders for helpful discussions. This work is
supported by BRNS grant Sanction No. 2007/37/60/BRNS/2913 dated 31.3.08,
Govt. of India.  
    
 

\end{document}